\documentstyle[preprint,eclepsf]{jpsj}
\def\runtitle{Strongly Correlated Electrons and Neutron Scattering}
\def\runauthor{Hidetoshi {\sc Fukuyama} }
\title{Strongly Correlated Electrons and Neutron Scattering}
\author{Hidetoshi Fukuyama\footnote{E-mail: fukuyama@phys.s.u-tokyo.ac.jp}}
\inst{Department of Physics, University of Tokyo,
Hongo, Bunkyo-ku, Tokyo 113-0033, Japan}
\abst{Various aspects of close and interesting relationship between
antiferromagnetism and singlet ground states are introduced for which
neutron scattering have been playing vital roles. The special emphasis is
on the disorder-induced antiferromagnetism in spin-Peierls systems, which
can be viewed as a nucleation process of classical magnetic order in the
background of singlet state whose origin is purely quantum-mechanical. It
is then pointed out that similar features will be found in Ce$_x$Cu$_2$Si$_2$ 
and high $T_{\rm c}$
cuprates. Finally the possible charge ordering process in NaV$_2$O$_5$ is
discussed which leads to the quenching of localized spins.}
\kword{A. Magnetic Materials, Superconductors, D. Magnetic Properties, 
Phase Transitions, Superconductivity}
\begin{document}
\sloppy
\maketitle

\section{Introduction}
Interactions between electrons in solids result in various states of
matter and phase transitions. Coulomb repulsive forces are known to be the
cause of magnetism. Especially in the presence of strong Coulomb
interaction, i.e. in strongly correlated systems, Mott insulators are
realized in the half-filled non-degenerate band, or in the case with
integer number of electrons per site in general, where the low lying
excitations are exclusively due to spin degrees of freedom. This situation
is properly described by the Heisenberg spin Hamiltonian. Here spins and
charges of electrons are completely separated. Although various types of
magnetic ground states have been explored in a transparent way by this
Heisenberg spin Hamiltonian, much of the recent interest are in the
nonmagnetic singlet ground states, which are usually termed  as
spin-gapped systems. These singlet ground states are purely
quantum-mechanical in their origin in comparison to magnetic states which 
are basically classical. Such studies on singlet ground states may be
motivated by the fact that the high temperature superconductivity in
cuprates, which has singlet $d_{x^2-y^2}$ symmetry, are realized next to 
AF Mott insulators, i.e. in the doped Mott insulators where
small amount of holes are introduced into the Mott insulating state \cite{1}.
Here it is to be noted that AF superexchange interaction, 
$J$, has two distinct aspects; it leads quite generally to
AF N\'eel ground states in lattices, while it results in the
singlet states obviously for two spins and even in lattices in the
presence of strong quantum fluctuations or frustrations. 
Actually it
turned out that there exists a close relationship between
antiferromagnetism and singlet $d_{x^2-y^2}$ superconductivity \cite{2}.  
In this paper some of the recent development of the studies on the
remarkable features of this interrelationship between antiferromagnetism 
and singlet ground states are introduced where neutron scattering
experiments have played decisive roles.

\section{Disordered Spin-Peierls Systems}
The spin-Peierls transition has been known for some time \cite{3,4} but the
discovery of this transition in inorganic compound, CuGeO$_3$, by Hase 
{\it et al}
\cite{5} is rather recent. 
Surprizing and puzzling was the report of the neutron
scattering experiment by Regnault {\it et al} \cite{6} on CuGeO$_3$, with a 
small amount of replacement of Ge by Si. 
This experiment has revealed the
coexistence of resolution-limited sharp Bragg spots, one corresponding to
the lattice dimerization stabilized below around the critical temperature
of spin-Peierls transition of the parent clean systems and the other to
AF ordering stabilized at much lower temperatures.
Since these two states, dimerized and AF states, had been considered to
be exclusive \cite{4}, some kind of inhomogeneity was suspected. 
At the same time, the  sharpness of these two Bragg spots together with 
the apparent transfer of the spectral intensity between these two spots as the
temperature is varied had suggested that the experiment had in fact
disclosed the coexistence of the true long range order of these two
conflicting order parameters. A theoretical proposal \cite{7} has been made,
which indicates that such coexistence is possible once the spacial
variations of the competing order parameters are taken into account. 
In this study the lattice distortions are treated completely classical because
of the three-dimensionality of the actual crystals and the quantum
mechanical features of one-dimensional spins are expressed in terms of the
phase Hamiltonian \cite{8} derived by the bosonization \cite{9}, which has been
treated in a mean field approximation in view of the existence of the
interchain exchange interaction. Basic physics behind turns out to be that
the realization of the singlet ground state by the dimerization is purely
quantum mechanical and the complete coherency of the wave functions of
spins in whole crystal (of the order of 10$^{23}$ !!) is needed  and that,
once the quantum phases of wave functions are perturbed, magnetizations
are created locally, which naturally order in the absence of frustrations. 
The excitations above this new type of ground state with the coexistence
turned out to have unique features \cite{10}: i.e. 
there are two distinct excitations, one with a gap reflecting
dimerization which is present even in the clean systems and the other at
very low energy and at around AF wavevector only  with
the very little total spectral weight in proportion to the degree of
disorder,i.e. impurity concentration. In the coexisting ground state this
small spectral weight at low energy forms well-defined spin wave mode
reflecting the long range AF order. This is schematically shown in Fig.1.
 
Such a feature of two-mode structure has actually been observed in neutron
scattering experiment \cite{11,12} and the existence of the well-defined spin
waves has also been proven by the ESR experiments as well\cite{13,14}. At
present the onset of the N\'eel ordering has been observed even down to
28.5mK at $x=5\times10^{-3}$ of Cu$_{1-x}$Zn$_x$GeO$_3$ by Manabe {\it et al}.
\cite{15}. Note that there is
little weight, i.e. ``transparent", in the energy region between the gap
and spin wave mode. This is very unusual and unexpected since the effects
of perturbations had been expected to be treated ``perturbatively" in the
presence of the gap in the excitation spectrum and then only the
modifications of the gapped modes could be expected. At the same time this
finding that the disorder introduces spectral weight only at very low
energy (as low as the elastic Bragg spots) even in the presence of large
gap will be a warning to correlate the experimental data of neutron
scattering and NMR without serious considerations on the sample quality,
since only the NMR will be affected by such low energy excitations.
The present problem of disorder-induced antiferromagnetism in spin-Peierls
systems can be considered to be a typical example showing how the spectral
weight of the excitations emerges in the process of nucleation. 
Here the first order transition is expected in the clean systems between
AF ordering and spin-Peierls dimerized state as a result of the competition 
between the interchain exchange interaction and the spin-lattice coupling
along the chain \cite{4,16}. Once the disorder is induced, however, the 
antiferromagnetism is nucleated around the impurities in otherwise singlet
ground state. 
\begin{figure}
\begin{center}
\epsfile{file=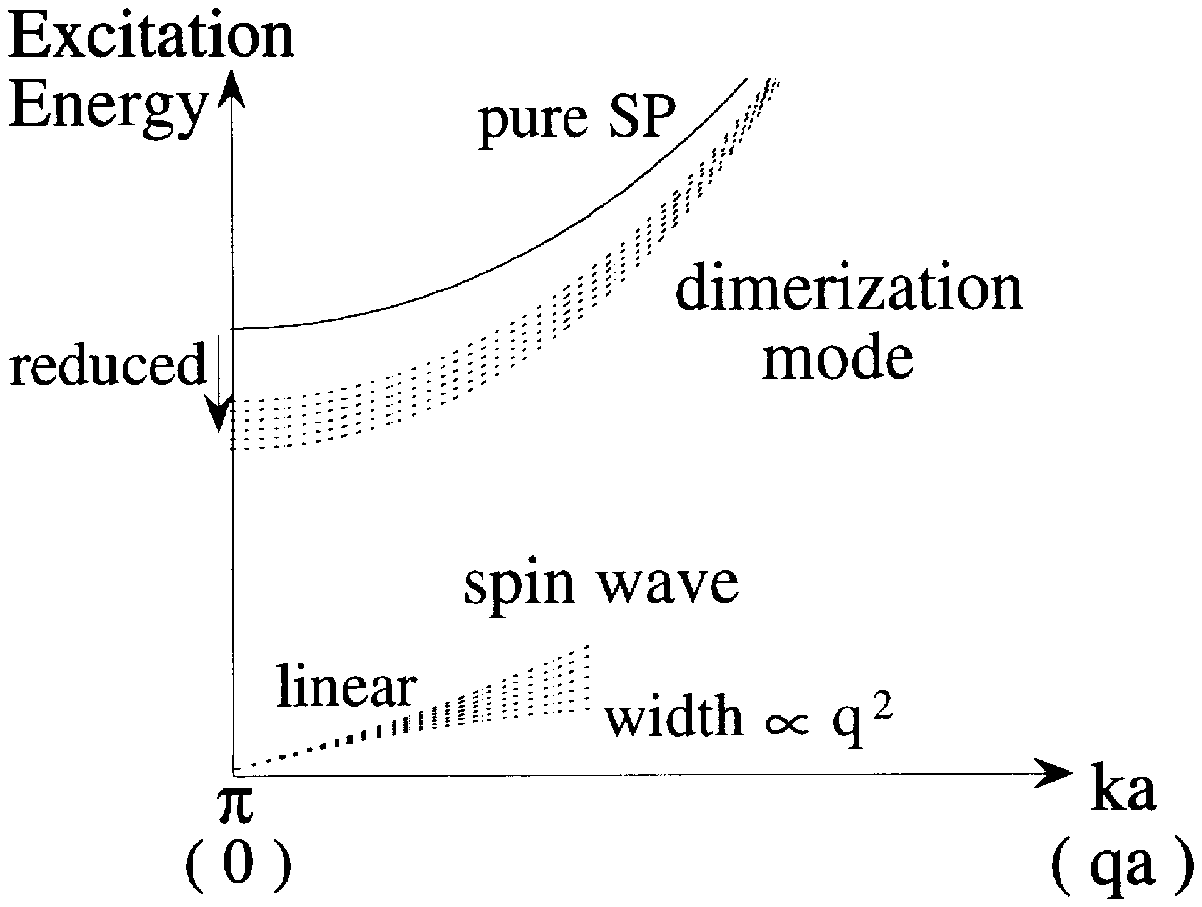,height=6cm}
\end{center}
\caption{A schematic representation of the distribution of the spectral weight
in the disordered spin-Peierls systems.
Note that disorder introduces small but finite amount of spectral weight at
very low energy only at AF wavevector.\cite{10}}
\end{figure}

\section{Heavy Electrons, Ce$_x$Cu$_2$Si$_2$}
Among strongly correlated electron sysytems, heavy electrons in lantanides
and actinides have been studied for a long time and it is now known that
magnetism and superconductivity are next to each other. Very recently
Ishida {\it et al} \cite{17}
found very mysterious properties in Cu-NQR spectra of series of 
Ce$_x$Cu$_2$Si$_2$.
At $x$=0.975 this system is AF ordered, while at $x$=1.025 the heavy electron
superconductivity is observed. However at intermediate $x$=0.99 they found
that there exist magnetic fluctuations with very low frequencies comparable
to the NQR frequency ($\omega$=10$^{6\sim7}$)Hz even though no signatures of
magnetism are seen in neutron scattering whose characteristic frequency is
$\omega$=10$^{11\sim12}$Hz. Since there should be intrinsic disorder due to 
the spacial
distribution of Ce atoms, a similar consideration as in the preceding
disordered spin-Peierls systems will apply and then a similar feature as in
Fig.1 
is expected though the gapped mode here is due to superconductivity
instead of dimerization in disordered spin-Peierls systems. 

\section{High $T_{\rm c}$ Cuprates}
Disclosures of the true nature of magnetic exciatations have played
crucial roles in the understanding of high $T_{\rm c}$ cuprates, for which both
neutron scaterring and NMR have contributed very much. Findings by the 
former will be fully discussed in this Symposium. On the other hand NMR 
experiment on the undredoped Y(123) by Yasuoka {\it et al}. \cite{18}
has first
indicated the existence of the spin gap: one of the most remarkable
features of the cuprates. All these experimental studies on magnetic
excitations have revealed that the disorder affects the excitations in
essential ways as in the case of disordered spin-Peierls systems. Few of
the examples will be discussed in the following. 
First the complete destructions of the spin-gap in NMR by a very small
amount of replacement of Cu by Zn in Y(248) \cite{19} 
will be understood again by Fig.1 if the spin-gap is associated with the
formation of the short range order of spin-singlet as described by the RVB
theory based on the slave boson mean field approximation of the $t$-$J$
model\cite{20} as shown in Fig.2 \cite{21}.
\begin{figure}
\begin{center}
\epsfile{file=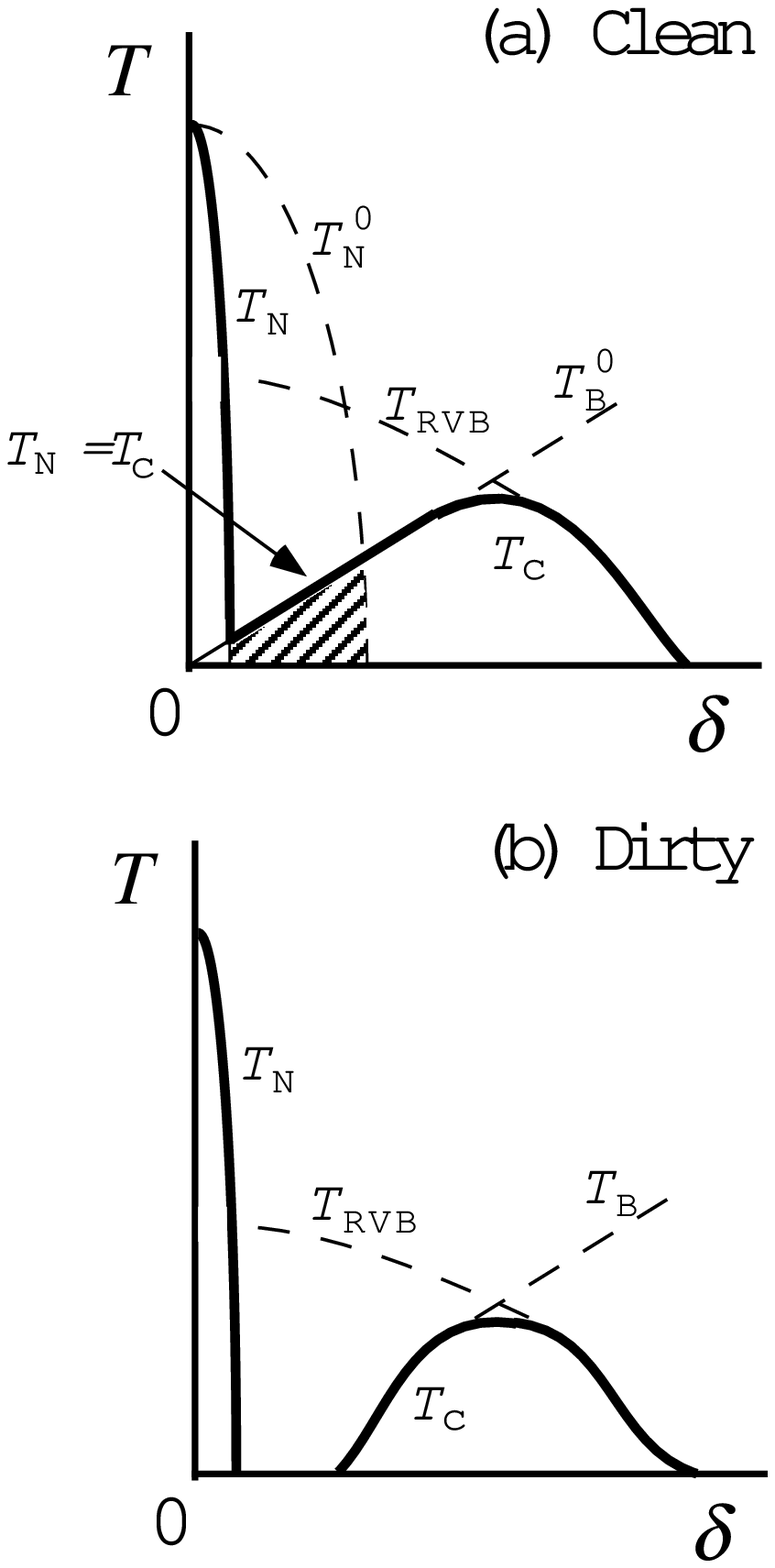,height=13cm}
\end{center}
\caption{A schematic representation of the phase diagram based on the slave
boson mean field theory of the $t$-$J$ model.
The spingap, which is the same as the pseudogap in this theory, is
associated with onset of singlet RVB order parameter.\cite{25}}
\end{figure}
Here only the general trends of the
spin-charge separation and their confinements \cite{22} are stressed. 
This spin gap can be identified with the pseudogap observed
later by ARPES experiments \cite{23,24} since electrons are convolutions of
spinons and holons \cite{24a}, as has been explicitly demonstrated \cite{25}. 
In this
context the underdoped region of La$_{2-x}$Sr$_x$CuO$_4$ is very interesting. 
Early neutron scattering data by Keimer {\it et al} \cite{26}
indicated the existence of spin
glass phase, which at the same time has a feature close to the long-range
ordered antiferromagnetgism of parent compounds. This finding has led the
present author to conjecture that the true long range AF ordering might
actually been realized at lower temperatures \cite{27}. However more recent
neutron studies \cite{28} unambiguously demonstrated that the ground state is a
canonical spin glass phase. A possible cause for this stability of
spin-glass over the long range AF ordered state may be due to the
extended nature of the wave function of the holes, which will lead to the
frustration among spins \cite{24}. This difference between extended and local 
disorder is obviously dependent on the mobility of holes, i.e. insulating
or (weakly) conducting, and may be related with the experimental findings
of shift of the spectral weight toward the low energy in accordance with 
the resistivity
increase as the temperature is lowered in underdoped YBCO \cite{29,30}.

\section{NaV$_2$O$_5$}
This system, which is insulating at all temperature, exhibits a magnetic
transition at $T_{\rm c}=35K$ below which the spin susceptibility drops 
sharply but
continuously \cite{31}, a feature typical of spin-gap systems, and has been
believed to be an example of spin-Peierls transition.In this case the
valences of V atoms are considered to be arranged as in Fig.3(b). 
\begin{figure}
\begin{center}
\epsfile{file=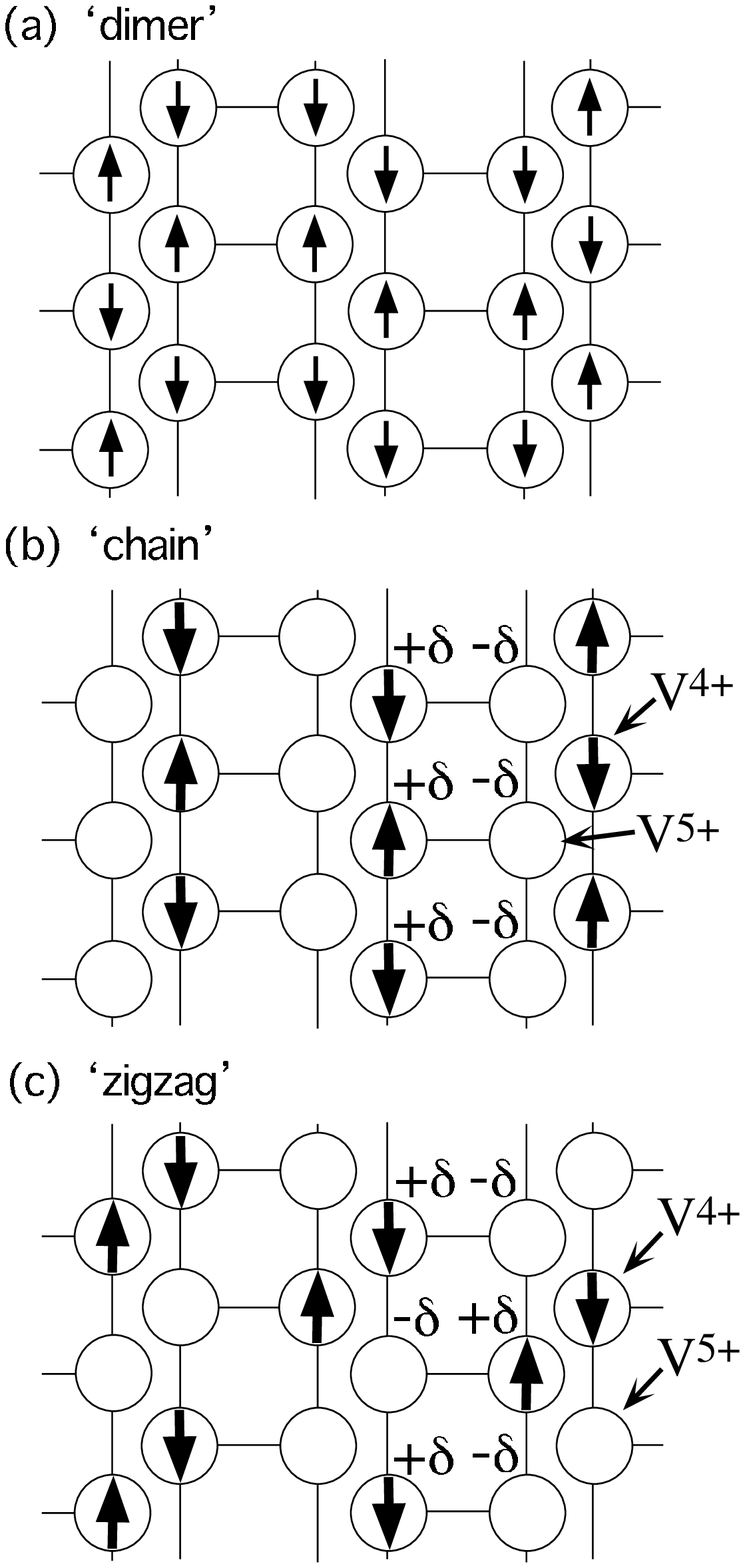,height=15cm}
\end{center}
\caption{A possible spacial pattern of charge ordering in NaV$_2$O$_5$.
\cite{33}}
\end{figure}
However
recent NMR experiment on V atoms by Ohama {\it et al} \cite{32} indicated 
that the
charge disproportionation sets in below $T_{\rm c}$ even though the the charge
distribution on V atoms is uniform above $T_{\rm c}$. 
This finding stimulated many
studies both theoretical and experimental. A theoretical study \cite{33} 
based on
the selfconsistent Hartree approximation for both onsite and intersite
Coulomb interaction has disclosed that the ground state will have a charge
ordering of zigzag type as seen in Fig.3(c), which has a spin dimer
structure. Above $T_{\rm c}$ the charge distributions are expected to be as in
Fig.3(a). Basically all experimental data appear to be consistent so far
with this state, including the spin excitation spectra probed by the
neutron scattering by Yosihama {\it et al} \cite{34,35}.  
It will be very interesting to probe the transient region of this charge
ordering which leads to the formation of the singlet ground state, i.e. the
quenching of the quantum spins.

\acknowledgements

The present research has been financially supported by the Grant-in Aid of
Monbushou on the Priority Area ``Anomalous Metallic State near the Mott
Transition" and is  based on the collaborations with
M. Saito, T. Tanimoto, N. Nagaosa, M. Sigrist, A. Furusaki for subject 2 
H. Kohno and T. Tanamoto for subject 4 and and H. Seo and H. Kino for subject 5 
whom I thank deeply. Discussions on experiments with
J.Akimitsu, Y. Endoh, Y. Fujii, K. Hirota, K. Kakurai, K. Katsumata, 
B. Keimer, N. Mori,
M. Motokawa,
M. Nishi, H. Nojiri, M. Sato, M. Sera, Z.X. Shen, G. Shirane, K. Yamada, 
H. Yasuoka 
K. Uchinokura
and Y. Ueda are fully appreciated. Special thanks are to Y. Kitaoka for
stimulating discussions on subject 3.

\end{document}